\documentclass[journal]{IEEEtran}
\usepackage{amsthm,amsmath,amssymb}
\usepackage{eucal}
\usepackage{xifthen}
\usepackage{mathtools}
\usepackage{enumerate}
\usepackage{microtype}
\usepackage{xspace}
\usepackage{bm}
\usepackage{cite}
\usepackage{fancyhdr}
\usepackage{lastpage}
\usepackage[small]{caption}
\usepackage{xcolor}
\allowdisplaybreaks

\long\def\comment#1{}

\newfont{\bbb}{msbm10 scaled 700}

\newfont{\bb}{msbm10 scaled 1100}

\renewcommand{\Re}[1][]{\ifthenelse{\isempty{#1}}{\operatorname{Re}}{\operatorname{Re}\left(#1\right)}}
\renewcommand{\Im}[1][]{\ifthenelse{\isempty{#1}}{\operatorname{Im}}{\operatorname{Im}\left(#1\right)}}

\newcommand{\CN}[1][]{\ifthenelse{\isempty{#1}}{\mathcal{N}_{\mathbb{C}}}{\mathcal{N}_{\mathbb{C}}\left(#1\right)}}
\renewcommand{\P}[1][]{\ifthenelse{\isempty{#1}}{\mathbb{P}}{\mathbb{P}\left(#1\right)}}
\newcommand{\E}[1][]{\ifthenelse{\isempty{#1}}{\mathbb{E}}{\mathbb{E}\left(#1\right)}}
\renewcommand{\det}[1][]{\ifthenelse{\isempty{#1}}{\mathrm{det}}{\mathrm{det}\left(#1\right)}}
\newcommand{\trace}[1][]{\ifthenelse{\isempty{#1}}{\mathrm{tr}}{\mathrm{tr}\left(#1\right)}}
\newcommand{\rank}[1][]{\ifthenelse{\isempty{#1}}{\mathrm{rank}}{\mathrm{rank}\left(#1\right)}}
\newcommand{\diag}[1][]{\ifthenelse{\isempty{#1}}{\mathrm{diag}}{\mathrm{diag}\left(#1\right)}}
\newcommand{\blkdiag}[1][]{\ifthenelse{\isempty{#1}}{\mathrm{blkdiag}}{\mathrm{blkdiag}\left(#1\right)}}

\renewcommand{\Re}{{\rm Re}}
\renewcommand{\Im}{{\rm Im}}

\allowdisplaybreaks

\DeclareMathAlphabet{\mathcal}{OMS}{cmsy}{m}{n}

\usepackage{algorithm}
\usepackage[algo2e]{algorithm2e} 
\usepackage{verbatim}
\usepackage{multirow}
\usepackage{url}
\usepackage{makecell}
\usepackage{float}
\usepackage{comment}
\usepackage{subfigure}
\usepackage{enumitem}
\usepackage{caption}

\usepackage{algpseudocode}
\usepackage[margin=0.7in]{geometry}
\newtheorem{Lemma}{Lemma}

\newtheorem{Definition}[Lemma]{Definition}

\newtheorem{Remark}{Remark}

  {\proof}{\proofend}
\newtheorem{Proposition}{Proposition}

\SetKwInput{KwInput}{Input}
\SetKwInput{KwOutput}{Output}

\begin{document}
\title{Positioning with Flexible Reflectors: Solution and Performance Analysis}
\author{Jiajun He, Han Yu, Hien Quoc Ngo, Michail Matthaiou
}
\author{
\IEEEauthorblockN{
Jiajun He\IEEEauthorrefmark{2}, Han Yu\IEEEauthorrefmark{1},
Danyan Lin\IEEEauthorrefmark{2}, 
Gaofeng Pan\IEEEauthorrefmark{4}, Hing Cheung So\IEEEauthorrefmark{3}, Stefano Buzzi\IEEEauthorrefmark{5}, and Hien Quoc Ngo\IEEEauthorrefmark{2}}
\thanks{The work of Han Yu was supported by the Alexander Humboldt Research Fellowship (grant No. 1240931).}
\thanks{The work of H. Q. Ngo was supported by a research grant from the Department for the Economy Northern Ireland under the US-Ireland R\&D Partnership Programme.}
%\IEEEauthorblockA{\IEEEauthorrefmark{2}E-mail: \{j.he, d.lin, hien.ngo\}@qub.ac.uk}

%\normalsize
\IEEEauthorblockA{\IEEEauthorrefmark{2}
       Centre for Wireless Innovation (CWI), Queen’s University Belfast, BT3 9DT Belfast, U.K.} 

\IEEEauthorblockA{\IEEEauthorrefmark{1}
        Faculty of Electrical Engineering and Computer Science, Technical University of Berlin,  Germany}
        
\IEEEauthorblockA{\IEEEauthorrefmark{4} School of Cyberspace Science and Technology, Beijing Institute of Technology, Beijing, China}

\IEEEauthorblockA{\IEEEauthorrefmark{3}Department of Electrical Engineering, City University of Hong Kong, Hong Kong SAR, China}

\IEEEauthorblockA{\IEEEauthorrefmark{5}Department of Electrical and Information Engineering, University of Cassino and Southern Lazio, Cassino, Italy}
}
\maketitle
\thispagestyle{empty}
\pagestyle{empty}

\begin{abstract}
Flexible reflectors (FRs) have emerged as a low-cost and energy-efficient solution for reshaping electromagnetic propagation environments across a wide range of applications. This paper investigates FR-swarm-assisted target localization in scenarios where line-of-sight (LoS) paths are unavailable. By leveraging the virtual LoS paths created by the FRs, a simple yet accurate estimator is proposed for localization under severe blockage conditions. To characterize the performance limits of the proposed scheme, we derive the Cramér-Rao lower bound (CRLB) and use it to optimize the positions and orientations of the FRs. Furthermore, by accounting for random FR deployment, we characterize the CRLB distribution and reveal how different network configurations affect localization accuracy. Simulation results demonstrate that the developed scheme closely approaches the CRLB performance, while the derived analytical results provide useful guidelines for FR deployment and network design.
\end{abstract}

%\begin{IEEEkeywords}
%Cramér-Rao lower bound, non-line-of-sight, received signal strength, reflection loss.
%\end{IEEEkeywords}

\vspace{-25pt}

\section{Introduction}

\IEEEPARstart{F}{uture} communication systems are expected to support seamless and ubiquitous services while meeting increasingly stringent key performance indicators (KPIs) \cite{Henk2026KPI}. Against this backdrop, metasurface technologies have attracted significant attention as promising solutions for enhancing network coverage, particularly in short-range communication scenarios, such as smart cities and industrial Internet of Things (IIoT). The devices usually operate under constraints on transmit power, hardware cost, and deployment complexity. In such scenarios, blockage, non-line-of-sight (NLoS) propagation, and rich multipath effects severely affect both link reliability and location awareness. A representative example is the intelligent reflecting surface (IRS) \cite{9326394}, which comprises multiple passive reflecting elements arranged over a two-dimensional (2-D) surface, with phase-control units used to steer incident signals toward target mobile devices. %Beyond conventional IRSs, semi-passive IRS has been proposed, where part of the surface is composed of passive reflecting elements and the remaining is equipped with active sensing elements to receive echo signals reflected from the target \cite{meng-irs}. In addition, \cite{mu2022star} introduced an advanced IRS architecture, namely the simultaneously transmitting and reflecting reconfigurable intelligent surface (STAR-RIS), which enables simultaneous signal transmission and reflection. 
Despite its advantages, the implementation cost of the RIS increases significantly with the number of reflecting elements, while accurate phase control remains difficult to achieve in practical deployments. Hence, a natural question arises: Is there a practical solution that can simultaneously reduce implementation cost and improve passive beamforming efficiency?

Fully passive metallic reflectors (PMR) have received considerable attention in recent years due to their low cost and implementation simplicity. Such reflectors are typically made of copper, aluminum, or conductive coatings in various shapes and have been widely used for coverage enhancement and channel-diversity improvement. The experimental results in \cite{rappaport2017indoor} demonstrate that a strong NLoS path can be created using PMR with received power approaching that of LoS propagation. In \cite{maeng2020coverage}, PMRs are employed to mitigate the severe blockage and propagation attenuation encountered in high-frequency bands, and the analytical results offer insights into how reflector placement and size can be designed to enhance coverage performance. Anjinappa $et$ $al.$ \cite{anjinappa2022placement} investigated PMR-assisted communications to enhance network coverage while reducing infrastructure cost. The results demonstrated that equivalent coverage performance can be attained with fewer BSs when a sufficient number of PMRs is deployed. To further improve the flexibility of PMR, \cite{lu2025FR} suggested a new architecture, namely the flexible reflector (FR), whose placement and orientation can be dynamically adjusted to optimize system performance. A mathematical framework for FR-based signal reflection was developed in \cite{10279522}, and its accuracy was verified through real-world experiments. Building upon \cite{lu2025FR}, Lu $et$ $al.$ presented a comprehensive investigation of passive beamforming in FR-aided communication systems through the joint optimization of FR positions and orientations. 

To our knowledge, PMR-assisted localization remains underexplored, while existing studies have mainly focused on communication-oriented tasks. Motivated by the unique advantages of FRs, this paper investigates FR-assisted positioning in scenarios where the LoS path is blocked. %To this end, a simple yet efficient localization algorithm is proposed for accurate target estimation. In addition, a comprehensive performance analysis is conducted, including both average and optimal localization performance in terms of the CRLB, to provide insights into how channel characteristics and FR deployment can be optimized to improve localization accuracy. 
The main contributions of this paper are summarized as follows.

\begin{enumerate}
    \item \textit{Low-Complexity Location Estimator}: We introduce an angle-dependent reflection model to characterize the FR-assisted channel. Based on this model, the FR-assisted positioning problem is formulated as a constrained weighted least-squares (CWLS) problem, which is solved via an iterative closed-form solution (CFS) despite its non-convex nature. Simulation results show that our algorithm closely approaches the optimal localization performance predicted by the CRLB.

    \item \textit{Optimum FR Placement and Optimal CRLB}: %In the absence of the LoS path, FRs enable first-order reflected paths for localization, and their deployment directly determines the achievable localization accuracy. 
    %To investigate the optimal system performance, 
    We derive the closed-form CRLB for FR-assisted localization and use it to investigate optimal FR placement. Since the resulting CRLB is lengthy and difficult to handle analytically, we further derive a simple approximation to the exact CRLB, based on which the optimal FR placement can be obtained without invoking numerical optimization methods. A simplified optimal CRLB is then derived to provide useful design guidelines for adjusting the placement and orientation of FRs toward optimal localization performance. 

    \item \textit{CRLB Distribution with Random FR Placements}: Conventional performance analyses of localization systems typically rely on fixed network geometries when deriving the CRLB. To overcome this limitation, we investigate the fundamental performance limits of FR-assisted positioning under random FR deployment and characterize how network configurations influence the overall localization performance in terms of the CRLB distribution.
\end{enumerate}

\begin{Remark}
    Due to the page limitation, the detailed derivations in the following analysis will be provided in the journal version.
\end{Remark}

\vspace{-0.5cm}

\begin{figure}[t]
\centerline{\includegraphics[width=0.6\columnwidth]{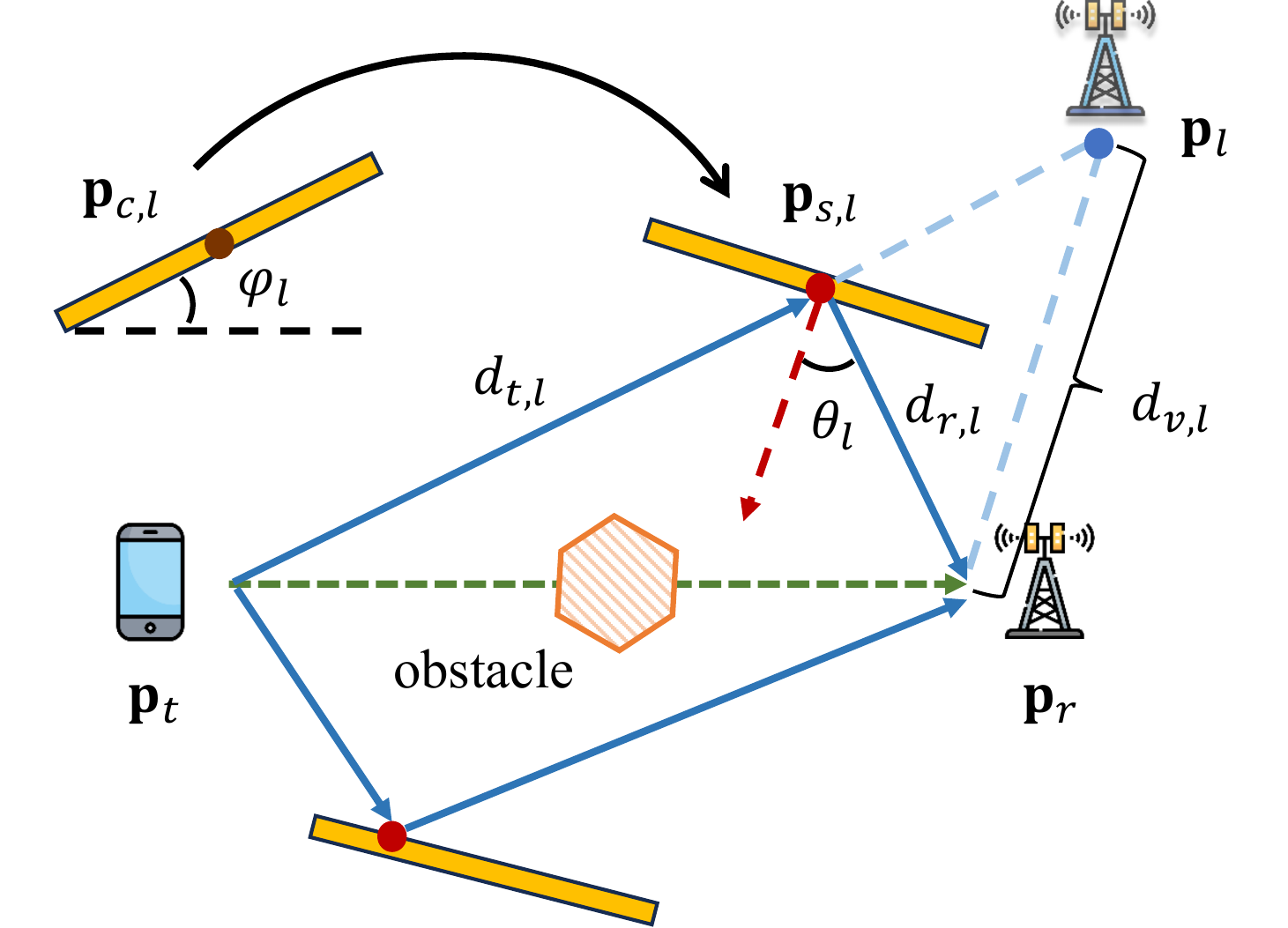}}
\caption{Configuration of the multi-FR target localization system.}
\label{fig-virtual link}
\vspace{-0.5cm}
\end{figure}

\section{System Model}

As shown in Fig. \ref{fig-virtual link}, we aim to estimate the target position\footnote{In this work, we consider a cooperative target that transmits pilot signals to the deployed Rx for localization.} $\mathbf{p}_t = [x_t \ y_t]^T$ in a 2-D space using received signal strength (RSS) measurements collected by a single receiver (Rx). Due to severe signal attenuation and blockage effects, particularly in high-frequency communication systems, the direct path between the target and the Rx is assumed to be blocked, such that localization relies solely on RSS samples obtained from FR-created reflection paths. The functionalities of the FR are provided as follows.

\begin{Definition}
    \textit{Flexible Reflector \cite{lu2025FR}}: The position and orientation of a FR can be configured to steer the reflection path toward the target, enabling passive beamforming control. Furthermore, the FR is equipped with on-off switching capabilities, enabling dynamic control over their reflection behavior.
\end{Definition}

\begin{Remark}
    During the localization process, the geometric information of the FRs is known \cite{han2017FR, lu2025FR}. In practice, all FRs can be connected to a central processing unit (CPU) to allow centralized control of their orientations and positions \cite{2026NCR}.    
\end{Remark}

\begin{Remark}
    Compared with conventional RISs, which require element-wise phase-shift optimization and control over a large number of reflecting units, FRs rely only on adjusting a few geometric variables, such as position and orientation \cite{lu2025FR}. Thus, FRs offer a lower-dimensional and simpler reconfiguration mechanism, which can substantially reduce reconfiguration latency and alleviate its impact on localization performance.
\end{Remark}

Given a Rx location at $\mathbf{p}_r = [x_r \ y_r]^T$, the received signal power from the reflection path between the target and Rx via the reflection point $\mathbf{p}_{s,l} = [x_{s,l} \ y_{s,l}]^T$ on the $l$-th FR is \cite{ho2026RSS}:
\begin{equation} \label{eq-received signal power}
 P_{r,l} = P_t G_{r} |\Gamma(\epsilon,\theta_l)|^2 (d_{t,l} + d_{r,l})^{-\alpha} w_l,
\end{equation}
where $P_t$ is the transmit power, $G_r$ is the antenna gain at the Rx, while $\alpha$ is the path-loss exponent; $w_l$ models RSS noise disturbance due to channel fading, and other unpredictable channel impairments, which is modeled by a log-normal random variable \cite{hcso-RSS}; $d_{t,l}$ and $d_{r,l}$ are distances from the reflection point on the $l$-th FR to the target and Rx, respectively, given by
$d_{t,l} = \| \mathbf{p}_t - \mathbf{p}_{s,l} \| , \; d_{r,l} = \| \mathbf{p}_r - \mathbf{p}_{s,l} \|.$
In addition, $\theta_l$ is the signal incident angle, given by
\begin{equation}
	\theta_l = \frac{1}{2}\arccos \left ( \frac{R^2 - d_{t,l}^2 - d_{r,l}^2}{2d_{t,l}d_{r,l}} \right ) , \; R = \|\mathbf{p}_t - \mathbf{p}_r \|. \label{eq-theta-o}	
\end{equation}	
The reflection gain is computed by: $|\Gamma(\epsilon,\theta_l)|^2 \in [0,1]$, which depends on both the dielectric permittivity $\epsilon$ of the reflecting surface and  $\theta_l$, as defined in \cite{permittivity}. Specifically, $|\Gamma(\epsilon, \theta_{l})| = a\theta_{l} + b$, where $a>0$ and $b>0$ are reflection coefficients that can be empirically determined via calibration and measurement campaigns \cite{hcso-RSS}. Given that the material of the FRs is known, the corresponding reflection coefficients can be determined.

Taking the logarithm on both sides of \eqref{eq-received signal power} and moving the terms related to the transmit power and antenna gain to the left-hand side, we have \cite{hcso-RSS}:
\begin{equation}
\begin{split}
    &\underbrace{\ln P_{r,l} -\ln P_t - \ln G_r}_{r_{\text{RSS},l}} \\
    &\triangleq \underbrace{-\alpha\ln (d_{t,l} + d_{r,l}) + 2\ln (a\theta_l + b)}_{r_{\text{RSS},l}^o \quad (\text{noise-free value})} + n_l, \\
\end{split}
\label{eq-nlos rss}
\end{equation}
where $n_l = \ln w_l$ is the additive white Gaussian noise (AWGN) with zero mean and variance $\sigma_n^2$ \cite{RSS-modeling}. 

\begin{Definition} Virtual Receiver (vRx):
\label{def-virtual anchor}
    Given a FR $l$ with center $\mathbf{p}_{c,l} = [x_{c,l} \ y_{c,l}]^T$ and inclination $\varphi_l$, the vRx is the mirror image of Rx with respect to (w.r.t.) the FR $l$ at position $\mathbf{p}_{l}$.
\end{Definition}

As a result, the coordinates of  vRx $l$, $\mathbf{p}_{l}$, are \cite{scatter}:
\begin{equation}
\label{eq-vRx}
\begin{split}
    x_{l} &= x_r - 2\tan\varphi_l\frac{\tan\varphi_l (x_r-x_{c,l}) - (y_r - y_{c,l})}{1+\tan^2\varphi_l}, \\
    y_{l} &=  y_r + 2\frac{\tan\varphi_l (x_r-x_{c,l}) - (y_r - y_{c,l})}{1+\tan^2\varphi_l}. \\
\end{split}
\end{equation}
By making use of \eqref{eq-vRx}, the RSS between the target and the Rx via the $l$-th reflection point is then given by:
\begin{equation}
    r_{\text{RSS},l} = -\alpha\ln d_l + 2\ln (a\theta_l + b) + n_l, 
\label{eq-RSS measurement}
\end{equation}
\begin{equation}
    \theta_l = \arccos \left ( \frac{R^2 - d_l^2 - d_{v,l}^2}{-2 d_l d_{v,l}} \right ),
\label{eq-incident angle}
\end{equation}
where $d_l = \| \mathbf{p}_{l} - \mathbf{p}_t \|$ and $d_{v,l} = \| \mathbf{p}_{l} - \mathbf{p}_r \|$. As indicated in \eqref{eq-nlos rss}, the reflection loss is modeled by the angle-dependent term $|\Gamma(\epsilon,\theta_l)|^2$ instead of the radar cross section. This angle-dependent term contains additional target-related information that can be leveraged to enhance localization performance. By properly adjusting the position and orientation of the FR, the specular reflection path can be guaranteed \cite{lu2025FR}. %while the corresponding reflection point on the reflector becomes the specular reflection point \cite{lu2025FR}. 
Assuming that $L$ FRs are used for target positioning, the RSS measurement vector is then given by:\footnote{The FRs remain fixed during the RSS acquisition phase, and their positions and orientations are subsequently adjusted based on the collected RSS samples.}
\begin{equation}
    \bm{r}_{\text{RSS}} = \bm{f}_{\text{RSS}}(\mathbf{p}_t) + \bm{n},
\label{eq-rss vector}
\end{equation}
\vspace{-20pt}
\begin{subequations}
\begin{align}
    \bm{r}_{\text{RSS}} &= 
    \begin{bmatrix}
        r_{\text{RSS},1}, & r_{\text{RSS},2}, & \ldots, & r_{\text{RSS}, L}
    \end{bmatrix}^T, \\
    \bm{f}_{\text{RSS}}(\mathbf{p}_t) &= 
    \begin{bmatrix}
        r^o_{\text{RSS},1}, & r^o_{\text{RSS},2}, & \ldots, & r^o_{\text{RSS}, L}
    \end{bmatrix}^T, \\
    \bm{n} &= 
    \begin{bmatrix}
        n_1, & n_2, & \ldots, & n_L
    \end{bmatrix}^T, 
\end{align}
\end{subequations}
while $r_{\text{RSS},l}^o$ denotes the noise-free RSS measurement between the target and the Rx via $l$-th reflection point. %Based on the RSS vector \eqref{eq-rss vector}, we propose a method to estimate the target position in the following section.

\vspace{-0.3cm}

\section{FR-assisted Target Positioning}

It is observed from \eqref{eq-RSS measurement} that the RSS signal model is highly nonlinear w.r.t. $\mathbf{p}_t$, which makes it difficult to obtain the global optimum. To address this challenge, an iterative algorithm is developed based on a sequence of convex approximations, thereby achieving high computational efficiency and robustness against poor initialization. %The proposed solution is initialized by establishing an approximate linear relationship between the collected RSSs and the target position, augmented with a set of auxiliary variables. 
Specifically, by applying a first-order Taylor expansion to $\theta_l$ in \eqref{eq-theta-o} around an appropriately chosen point $\kappa$ and using $\mathbf p_l$, we obtain
\begin{align}\label{t2}
    \theta_l &\simeq \arccos\kappa - \frac{\left(\frac{(\mathbf{p}_{l}-\mathbf{p}_r)^T(\mathbf{p}_{l}-\mathbf{p}_t)}{ d_l d_{v,l}}-\kappa\right)}{\sqrt{1-\kappa^2}}.
\end{align}
By rearranging \eqref{eq-RSS measurement} and moving the term $-\alpha \ln d_l + n_l$ to the left-hand side, exponentiating both sides, and applying the approximation in \eqref{t2}, we have
\begin{align}
\label{rr2}
    &\quad d_l^{\alpha/2} \exp \left ( \frac{r_{\text{RSS},l}-n_l}{2} \right )  \simeq  a\arccos\kappa+ \frac{a\kappa}{\sqrt{1-\kappa^2}}+b \nonumber \\
    &\qquad \qquad  - \frac{a}{d_{v,l}\sqrt{1-\kappa^2} } \frac{(\mathbf{p}_{l}-\mathbf{p}_r)^T(\mathbf{p}_{l}-\mathbf{p}_t)}{ d_l }.
\end{align}
Multiplying both sides of \eqref{rr2} by $d_l$ and performing a first-order Taylor expansion on $n_l$ around zero leads to
\begin{equation}
    \begin{split} 
\label{t4}
    &d_l^{1+\alpha/2} e^{r_{\text{RSS},l}/2} - \left (a\arccos\kappa+ \frac{a\kappa}{\sqrt{1-\kappa^2}}+b \right ) d_l  \\
    &+ \frac{a (\mathbf{p}_{l}-\mathbf{p}_r)^T(\mathbf{p}_{l}-\mathbf{p}_t)}{ d_{v,l}\sqrt{1-\kappa^2}} \simeq  d_l^{1+\alpha/2}e^{r_{\text{RSS},l}/2}/2  n_l .
  \end{split}
\end{equation}
By defining $\bm y^o =[\mathbf{p}_t^T, d_1, \ldots,d_L,d_1^{1+\alpha/2},\ldots,d_L^{1+\alpha/2}]^T$, 
% \begin{align}
%     \bm y^o =[\mathbf{p}_t^T, d_1, \ldots,d_L,d_1^{1+\alpha/2},\ldots,d_L^{1+\alpha/2}]^T,
% \end{align}
%\eqref{t4} for $l=1,2,\ldots,L$ can be stacked into a more concise matrix form as
we have the following
\begin{align}
\label{t5}
    \bm A\bm y^o+\bm h \simeq \bm B \bm n,
\end{align}
where the $l$-th row of matrix $\bm A$ is defined by
\begin{equation}
    \begin{split}
    \bm A_{(l,:)} &= \bigg[-\frac{a}{ d_{v,l}\sqrt{1-\kappa^2}} (\mathbf{p}_{l}-\mathbf{p}_r)^T , \bm 0_{1\times(l-1)}, \\
    &\qquad- \left ( a\arccos\kappa + \frac{a\kappa}{\sqrt{1-\kappa^2}}+b \right ), \\
    &\qquad \bm 0_{1\times(L-1)}, e^{r_{\text{RSS},l}/2},\bm 0_{1\times(L-l)}\bigg],        
    \end{split}
\end{equation}
and the $l$-th element of vector $\bm h$ is defined by $h_{(l)} = \frac{a}{ d_{v,l}\sqrt{1-\kappa^2}}  (\mathbf{p}_{l}-\mathbf{p}_r)^T\mathbf{p}_{l}$, 
% \begin{align*}
%     h_{(l)} = \frac{a}{ d_{v,l}\sqrt{1-\kappa^2}}  (\mathbf{p}_{l}-\mathbf{p}_r)^T\mathbf{p}_{l},
% \end{align*}
while $\bm B$ is a diagonal matrix with the $l$-th diagonal element being $B_{(l,l)} = d_l^{1+\alpha/2}e^{r_{\text{RSS},l}/2}/2$. Based on \eqref{t5}, a non-convex CWLS problem can then be: 
\begin{equation}
    \begin{split}
        \label{cwls1}
	\min_{\bm y}\;\;&(\bm A\bm y+\bm h )^{T}\bm W(\bm A\bm y+\bm h) \\
	{\rm{s.t.}}\;\;&y_{(2+l)} = \|\bm y_{(1:2)}-\mathbf p_{l}\|, \\
	&y_{(2+N+l)} = y_{(2+l)}^{1+\alpha/2},\ l=1,\ldots,N,
\end{split}
\end{equation}
where $\bm y$ is the optimization variable corresponding to $\bm y^o$, 
    $\bm W=\mathbb{E} \left \{ (\bm B \bm n\bm n^T \bm B ^T)^{-1} \right \} =(\bm B \bm Q \bm B ^T)^{-1}$ 
is the weighting matrix with $\bm Q$ being the covariance matrix of $\bm n$. Since the constraints in \eqref{cwls1} are non-convex, the resulting problem cannot be solved efficiently in closed form. We first obtain an initial solution by relaxing these constraints, i.e.,
\begin{align}\label{y1}
    \hat{\bm y} = -(\bm A^T\bm W\bm A)^{\dagger}\bm A^T\bm W\bm h,
\end{align}
where $(\cdot)^{\dagger}$ is the Moore-Penrose inverse. Thus, a coarse estimate of the target position, denoted $\hat{\mathbf{p}}_t$, is then extracted from the first two elements of $\hat{\bm y}$.

\begin{algorithm}[t]
\small
\centering
\begin{tabular}{@{}p{\linewidth}@{}}
        \textbf{Algorithm 1:} Multi-FR-assisted localization algorithm \\
        \hline
        \textbf{Input}:\\
        \qquad $\bm r_{\rm RSS}$: RSS samples;\ \{$a,b$\}: Reflection coefficients;\\
        \qquad $\bm Q$: Covariance matrix of measurement noise;\\
        \qquad$\kappa=0, i=0$: Initialize $\kappa$ and $i$;\ $\bm W=\bm I$;\\ %Initialize $\bm W$ to the identity matrix;\\
        \textbf{Steps}:\\
        \qquad 0: Obtain initial target estimate by using \eqref{y1} \\
        \qquad \quad and set it as $\mathbf{p}_t^0$; \\
        \qquad 1: Substitute $\mathbf{p}_t^0$ into \eqref{mlc} and let $C^0=C({}_{t}^0)$; \\
        \qquad 2: Update $\kappa$, $\bm A$,  $\bm W$, and $\bm G$ by $\mathbf p_t^0$; \\
        \quad \textbf{do while}\;$i<i_{\rm max}$\\
        \qquad 3: $i=i+1$; \\
        \qquad 4: Obtain $\mathbf p_t^i$ by solving \eqref{yhat}, and calculate \\
        \qquad\quad $C^i=C({\mathbf{p}}_{t}^i)$ using \eqref{mlc}; \\
        \qquad \textbf{if}\; $C^{i}-C^{i-1}>\delta_c$ \\
        \qquad\qquad 5: Set $i=0$ and randomly generate a new $\mathbf p_t^0$, \\
        \qquad\qquad\quad and update $C^0$;\\
        \qquad \textbf{end} \\
        \qquad \textbf{if}\; $i>0$ \text{and} $\|\mathbf p_t^i-\mathbf p_t^{i-1}\|<\delta_p$, \\
        \qquad\quad \textbf{break}; \\
        \qquad 6: Update $\kappa$, $\bm A$, $\bm W$, and $\bm G$ using ${\mathbf{p}}_{t}^i$.\\
        \quad \textbf{end}\\
        \textbf{Output}: The target position estimate \\
\end{tabular}
\end{algorithm}

To refine this estimate, we linearize the equality constraints around $\hat{\bm y}$ using the first-order Taylor expansion. The linearized constraints can be expressed in a compact matrix form as
\begin{align}
    \bm G \bm y + \bm g =\bm 0,
\label{eq:linearized_constraints}
\end{align}
\vspace{-0.6cm}
\begin{align}
    &\bm G_{(l,:)} = \left [-\frac{\hat {\bm y}_{(1:2)}^T-{\bf{p}}_{l}^T}{\|\hat {\bm y}_{(1:2)}-{\bf p}_{l}\|},\bm 0_{1\times(l-1)},1,\bm 0_{1\times(2N-l)} \right ], \nonumber\\
    &\bm G_{(N+l,:)} = \left [ - (1+\alpha/2)\hat y_{(2+l)}^{\alpha/2-1},\bm 0_{1\times(N+l-1)},1,\bm 0_{1\times(N-l)} \right ], \nonumber\\
    & g_{(l)} = -\|\hat {\bm y}_{(1:2)}-\bm p_{l}\|+\frac{\hat {\bm y}_{(1:2)}^T-\bm p_{l}^T}{\|\hat {\bm y}_{(1:2)}-\bm p_{l}\|} \hat {\bm y}_{(1:2)},\nonumber\\
    & g_{(N+l)}=- \hat y_{(2+l)}^{1+\alpha/2} + (1+\alpha/2)\hat y_{(2+l)}^{\alpha/2-1}\hat y_{(2+l)}. \nonumber
\end{align}
The substitution of \eqref{eq:linearized_constraints} into \eqref{cwls1} converts the original non-convex problem into the following convex quadratically constrained linear program:
\begin{equation}
    \begin{split}
\label{cwls2}
	\min_{\bm y}\;\;&(\bm A\bm y+\bm h )^{T}\bm W(\bm A\bm y+\bm h) \\
	{\rm{s.t.}}\;\;&\bm G \bm y + \bm g =\bm 0,
 \end{split}
\end{equation}
while the optimal solution for \eqref{cwls2} can be obtained efficiently by solving its Karush-Kuhn-Tucker (KKT) conditions, leading to the following closed-form update:
\begin{align}\label{yhat}
    \begin{bmatrix}
        \tilde {\bm y} \\
        \tilde {\bm \lambda}
    \end{bmatrix} = -\begin{bmatrix}
        2\bm A^T\bm W\bm A &\bm G^T\\
        \bm G &\bm 0
    \end{bmatrix}^{-1} \begin{bmatrix}
        2\bm A^T \bm W \bm h \\
        \bm g
    \end{bmatrix},
\end{align}
where $\bm{\lambda}$ is the Lagrange multiplier vector. This update is performed iteratively. At each iteration, the expansion point $\kappa$, along with $\bm{A}$, $\bm{W}$, and $\bm{G}$, is recalculated based on the current estimate $\hat{\bm{y}}$. Nevertheless, a critical challenge is that the initial unconstrained solution obtained from \eqref{y1} may be unstable due to the ill-posed nature of the problem, while such instability can prevent the subsequent iterative procedure from converging. Thus, a re-initialization strategy is employed. When the iterative process fails to converge, e.g., when the cost function does not decrease monotonically, the algorithm is re-initialized with a random starting point. Such a random initialization is accepted only if it leads to successful convergence, while the final solution is selected as the one yielding the minimum maximum likelihood (ML) cost function value, i.e.,
\begin{align}\label{mlc}
     C({\mathbf{p}}_{t} ) = (\bm{r}_{\text{RSS}} - \bm{f}_{\text{RSS}}({\mathbf{p}}_{t} ))^T \bm Q^{-1} (\bm{r}_{\text{RSS}} - \bm{f}_{\text{RSS}}({\mathbf{p}}_{t} )). 
\end{align}
The complete procedure is summarized in \textbf{Algorithm 1}.

\vspace{-0.2cm}

\section{Performance Analysis}

\subsection{Target Observability}

Target observability originates from control theory \cite{Montanari2023ob} and characterizes the theoretical possibility of uniquely determining the target position from noise-free measurements. %Target observability has recently attracted considerable attention and has been widely used to analyze the performance of positioning systems. 
Before assessing the feasibility of the FR-aided localization scheme, the Jacobian matrix of the RSS signal model, i.e., $\bm{J}_{\mathbf{p}_t} = \frac{\partial \bm{f}_{\text{RSS}}(\mathbf{p}_t)}{\partial \mathbf{p}_t}$, is given by the following lemma.

\begin{Lemma}
\label{lemma-Jacobian}
    Assuming that the LoS path is blocked and that $L$ FRs are employed for localization, the Jacobian matrix of the RSS vector in \eqref{eq-rss vector} is given by
    \begin{equation}
    \begin{split}
        &\bm{J}_{\mathbf{p}_t} = \frac{\partial \bm{f}_{\text{RSS}}(\mathbf{p}_t)}{\partial \mathbf{p}_t}= \\
        &\!\!\!\!\! 
        \begin{bmatrix}
            \bar{\alpha}_{1}(x_t - x_{1}) + \bar{\beta}_1 (x_{1}-x_r) & \!\!\!\!\!\!\!\bar{\alpha}_{1}(y_t - y_{1}) + \bar{\beta}_1 (y_{1}-y_r) \\
            \vdots & \!\!\!\!\!\!\! \vdots \\
            \bar{\alpha}_{L}(x_t - x_{L}) + \bar{\beta}_L (x_{L}-x_r) & \!\!\!\!\! \bar{\alpha}_{L}(y_t - y_{L}) + \bar{\beta}_L (y_{L}-y_r) \\
        \end{bmatrix},
    \end{split}
    \label{eq-J closed-form}
    \end{equation}
    \vspace{-0.5cm}
    \begin{align}
        \bar{\alpha}_l &= -\frac{\alpha}{d_l^2} + \delta_l \frac{(R^2 - d_l^2 - d_{v,l}^2)}{2 d_{v,l} d_l^3}, \\
        \bar{\beta}_l &= \frac{\delta_l}{d_{v,l} d_l} \
        \delta_l = \frac{2a}{a\theta_l+b}\frac{-1}{\sqrt{1-\bar{\theta}^2}}.
    \end{align}
\end{Lemma}

Based on \textbf{Lemma \ref{lemma-Jacobian}}, the necessary and sufficient conditions required to satisfy the target observability are:

\begin{enumerate}
    \item \textit{Necessary Condition}: A fundamental requirement for target observability is that the number of independent RSS measurements shall be no smaller than the number of unknown parameters to be estimated. For a 2-D target position, this condition requires that $L \geq 2$. Therefore, \textit{at least} two FRs providing distinct NLoS paths are required.

    \item \textit{Sufficient Condition}: A sufficient condition for achieving the target observability is that the FIM is invertible for the proposed localization problem, which is equivalent to the Jacobian matrix $\bm{J}_{\mathbf{p}_t}$ being full column rank.
\end{enumerate}

Based on the above conditions, we further analyze the determinant of $\bm{J}_{\mathbf{p}_t}$ for the minimal case of $L=2$ to obtain geometric insights. In this case, the determinant expression can be expanded into a sum of four terms:
\begin{equation}
    \begin{split}
    &|\bm{J}_{\mathbf{p}_t}| = \left [\bar{\alpha}_{1}(x_t - x_{1}) + \bar{\beta}_1 (x_{1}-x_r) \right ] \\
    &\qquad\quad \ \left [\bar{\alpha}_{2}(y_t - y_{2}) + \bar{\beta}_2 (y_{2}-y_r) \right ] \\
    &\qquad \ -[\bar{\alpha}_{2}(x_t - x_{2}) + \bar{\beta}_2 (x_{2}-x_r)] \\
    &\qquad\quad \ \ [\bar{\alpha}_{1}(y_t - y_{1}) + \bar{\beta}_1 (y_{1}-y_r)] \\
    &=\bar{\alpha}_{1}\bar{\alpha}_{2}|\mathbf p_t-\mathbf p_{1},\mathbf p_t-\mathbf p_{2}|+ \bar{\alpha}_{1} \bar{\beta}_2 |\mathbf p_t-\mathbf p_{1},\mathbf p_{1}-\mathbf p_r| \\
    &+\bar{\alpha}_{2}\bar{\beta}_{1}|\mathbf p_{1}-\mathbf p_r,\mathbf p_t-\mathbf p_{2}| + \bar{\beta}_{1} \bar{\beta}_2 |\mathbf p_{1}-\mathbf p_r,\mathbf p_{2}-\mathbf p_r|,   
    \end{split}
\end{equation}
where the scalar coefficients $\bar{\alpha}_l$ and $\bar{\beta}_l$ are functions of the path lengths and incident angles, which are typically nonzero. Consequently, the Jacobian matrix $\bm{J}_{\mathbf{p}_t}$ is of full rank if at least one of these four determinant terms is nonzero. This yields the following sufficient geometric conditions:
\begin{itemize}
    \item \textit{Non-collinearity with the target}: The two reflectors are not collinear with the target, which ensures that the vectors $(\mathbf{p}_t - \mathbf{p}_{1})$ and $(\mathbf{p}_t - \mathbf{p}_{2})$ are linearly independent. Consequently, the first determinant term is nonzero.
    
    \item \textit{Non-collinearity with the Rx}: The two reflectors are not collinear with the Rx, making the last term nonzero, as the vectors $(\mathbf{p}_{1} - \mathbf{p}_r)$ and $(\mathbf{p}_{2} - \mathbf{p}_r)$ are independent.
    
    \item \textit{Non-collinearity in target-reflector-Rx geometry}: At least one reflector is not collinear with both the target and the Rx. This ensures that the mixed determinant terms, e.g., those involving $(\mathbf{p}_t - \mathbf{p}_{1})$ and $(\mathbf{p}_{1} - \mathbf{p}_r)$, are nonzero.
\end{itemize}
In summary, the system is observable with $L=2$ provided that the two FRs are repositioned to avoid the aforementioned degenerate collinear configurations. For $L>2$, observability is guaranteed provided that at least two reflectors satisfy the above non-linearity conditions.

\vspace{-0.38cm}

\subsection{Optimal CRLB}

%The CRLB is commonly adopted to evaluate the localization performance of an unbiased estimator. 
For the considered FR-aided localization problem, we have
\begin{equation}
    \text{CRLB}(\bm{r}_{\text{RSS}}) \triangleq \text{tr} \left ( \text{FIM}^{-1}(\bm{r}_{\text{RSS}}) \right ),
\label{eq-CRLB exact}
\end{equation}
\begin{equation}
\begin{split}
   \text{FIM}(\bm{r}_{\text{RSS}}) 
   &= \left [\frac{\partial \bm{f}_{\text{RSS}}(\mathbf{p}_t)}{\partial\mathbf{p}_t} \right ]^T \bm{C}_{\text{RSS}}^{-1} \left [ \frac{\partial \bm{f}_{\text{RSS}}(\mathbf{p}_t)} {\partial\mathbf{p}_t} \right ]. \\
\end{split}
\label{eq-FIM exact}
\end{equation}
% The Jacobian $\bm{J}_{\mathbf{p}_t} = {\partial \bm{f}_{\text{RSS}}(\mathbf{p}_t)}/{\partial \mathbf{p}_t}$ is computed according to \textbf{Lemma \ref{lemma-Jacobian}}. By substituting \eqref{eq-J closed-form} and \eqref{eq-FIM exact} into \eqref{eq-CRLB exact} then yields
% \begin{equation}
% \text{tr} \left ( \text{CRLB}(\bm{r}_{\text{RSS}}) \right ) = \sigma_n^2\frac{(\vartheta_1 + \vartheta_2)}{\vartheta_1\vartheta_2 - \vartheta_3^2},
% \label{eq-crb closed-form}
% \vspace{-0.2cm}
% \end{equation}
% {\begin{align}
%     \vartheta_1 & = \sum_{l=1}^{L} \left ( 
%     \begin{bmatrix}
%         \bar{\alpha}_l \\ \bar{\beta}_l 
%     \end{bmatrix}^{'} 
%     \begin{bmatrix}
%         x_t - x_l \\ x_l - x_r
%     \end{bmatrix}
%     \right )^2, \nonumber \\
%     \vartheta_2  &= \sum_{l=1}^{L} \left ( 
%     \begin{bmatrix}
%         \bar{\alpha}_l \\ \bar{\beta}_l 
%     \end{bmatrix}^{'}
%     \begin{bmatrix}
%         y_t - y_l \\ y_l - y_r
%     \end{bmatrix} 
%     \right )^2, \nonumber \\
%     \vartheta_3 &= \sum_{l=1}^{L} 
%     \left ( 
%     \begin{bmatrix}
%         \bar{\alpha}_l \\ \bar{\beta}_l 
%     \end{bmatrix}^{'} 
%     \begin{bmatrix}
%         x_t - x_l \\ x_l - x_r
%     \end{bmatrix}
%     \right )
%     \left ( 
%     \begin{bmatrix}
%         \bar{\alpha}_l \\ \bar{\beta}_l 
%     \end{bmatrix}^{'}
%     \begin{bmatrix}
%         y_t - y_l \\ y_l - y_r
%     \end{bmatrix} 
%     \right ),   
% \label{FIM-co1}
% \end{align}
% where $(x_l, y_l)$ is the vRx position given by \eqref{eq-vRx}, while $\bar{\alpha}_{l}$ and $\bar{\beta}_{l}$ are in \textbf{Lemma \ref{lemma-Jacobian}}.
Nevertheless, directly minimizing the exact CRLB in \eqref{eq-CRLB exact} using a genetic algorithm (GA) provides limited analytical insight into the optimal placement of FRs. Thus, we provide a tractable approximation of tr(CRLB) in the following.

\begin{Proposition}
By assuming that $L$ FRs are used for the target positioning, the value tr(CRLB) for the FR-assisted localization problem can be approximated by:
\begin{subequations}\label{eq-ABC-CRLB}
	\begin{gather}
    \text{tr}(\text{CRLB}(\bm{r}_{\text{RSS}})) \approx \frac{\sigma_n^2(\tilde{\vartheta}_1+\tilde{\vartheta}_2)}{\tilde{\vartheta}_1 \tilde{\vartheta}_2 - \tilde{\vartheta}_3^2}, \\
    \tilde{\vartheta}_1 = \sum_{l=1}^{L} \left ( \alpha + \frac{2a}{a\theta_{l}+b} \right )^2 \frac{2\theta_l^2}{R^2} \cos^2\phi_l, \label{eq-case 1 A} \\
     \tilde{\vartheta}_2 = \sum_{l=1}^{L} \left ( \alpha + \frac{2a}{a\theta_{l}+b} \right )^2 \frac{2\theta_l^2}{R^2} \sin^2\phi_l, \label{eq-case 1 B} \\
    \tilde{\vartheta}_3 = \sum_{l=1}^{L} \left ( \alpha + \frac{2a}{a\theta_{l}+b} \right )^2 \frac{2\theta_l^2}{R^2} \sin\phi_l\cos\phi_l, \\
  \cos\phi_l = (x_t - x_l)/d_l \, , \;\; \sin\phi_l = (y_t - y_l)/d_l.  \label{appr-1}  
    \end{gather}
\end{subequations}
\end{Proposition}

By optimizing the positions and orientations of the FRs using \eqref{eq-ABC-CRLB}, we obtain the following proposition.

\begin{Proposition}
\label{prop-lower bound}
To consider the critical value $\{ \theta_l \}_{l=1}^{L} =  90^{\circ}$, the tight lower bound of the optimal CRLB, $\text{CRLB}(\bm{r}_{\text{RSS}})_{o}$, is:
\begin{equation}
     {\text{tr}\left ( \text{CRLB}(\bm{r}_{\text{RSS}})_{o} \right ) %= \sigma_n^2\frac{R^2}{L\cdot(\frac{\pi}{2}\alpha+\frac{4a\pi}{a\pi+2b})^2},
     > \sigma_n^2\frac{R^2}{2L (\pi\alpha+4)^2}}.
\label{eq-optimal CRLB case 1}
\end{equation}  
%where $R$ is the distance between the target and the Rx. 
The optimized geometry ensures $\vartheta_3 = 0$.
\end{Proposition}

%It is worth mentioning that the minimum achievable distance is bounded by the reference distance $R_{\text{ref}}$ for RSS calibration, since the RSS remains a constant value when $R < d_{\text{ref}}$.

\vspace{-0.45cm}

\subsection{CRLB Distribution}

Furthermore, we compare the optimal localization performance given by \textbf{Proposition \ref{prop-lower bound}} with the average CRLB in order to quantify the performance gain obtained through optimized FR placement. To compute the average CRLB, the randomness of the FRs is introduced using the Boolean line model (BLM), which has been widely adopted to simulate the random deployment of the RIS. In our study, a FR $l$ is characterized as a line segment with its center point $\mathbf{p}_{c,l}$, length $\ell_l$, and inclination angle $\varphi_l$. We denote each FR by $\mathcal{R}_{\mathbf{p}_{c,l}, \ell_l, \varphi_l}$. The center point $\mathbf{p}_{c,l}$ is modeled by the Homogeneous Poisson Point Process (HPPP) with density $\lambda_r$ within the disk $\bm{b}(\textit{O},R_a)$. The FR length $\ell_l$ is generated from the uniform distribution $\mathcal{U}[\ell_{\text{min}},\ell_{\text{max}}]$. The inclination angle $\varphi_l$ and the orientation angle $\Phi_l$ are independent, and they are sampled uniformly from $[0,\pi]$ rad. In this setting, the average CRLB can be computed using the following lemma.

% \begin{figure*}[!t]
% \normalsize
% %\setcounter{MYtempeqncnt}{\value{equation}}
%   %\setcounter{equation}{1}
% \begin{equation}
% \begin{split}
%     \text{CRLB}(\bm{r}_{\text{RSS}}) %\approx \sigma_n^2\cdot\frac{\vartheta_1+\vartheta_2}{\vartheta_1\vartheta_2} \\
%     &\approx \frac{\sigma_n^2 \left (\sum_{l=1}^{L} \left ( \alpha + \frac{2a}{a\theta_{l}+b} \right )^2 \frac{2\theta_l^2}{R^2} (\sin^2\phi_l + \cos^2\phi_l\right )}{\sum_{l=1}^{L} \left ( \alpha + \frac{2a}{a\theta_{l}+b} \right )^2 \frac{2\theta_l^2}{R^2} \cos^2\phi_l \times \sum_{l=1}^{L} \left ( \alpha + \frac{2a}{a\theta_{l}+b} \right )^2 \frac{2\theta_l^2}{R^2} \sin^2\phi_l} \\
%     &\approx \frac{\sigma_n^2 L R^2}{2\sum_{l=1}^{L} \left ( \alpha + \frac{2a}{a\theta_{l}+b} \right )^2 \theta_l^2 \cdot \cos^2\phi_l \sum_{l=1}^{L} \sin^2\phi_l} \\
%     &\overset{(a)}{\approx} \frac{\sigma_n^2 L R^2}{2\sum_{l=1}^{L} a_l^2 b_l^2 \sum_{l=1}^{L} 1-b_l^2} \overset{(b)}{\approx} \frac{\sigma_n^2 L^2 R^2}{2\sum_{l=1}^{L} a_l^2 \left (L\sum_{l=1}^{L} b_l^2 -\left (\sum_{l=1}^{L}b_l^2 \right )^2 \right )}, \\
% \end{split}
% \label{eq-crb mid}
% \end{equation}
% %\setcounter{equation}{\value{MYtempeqncnt}}
% \hrulefill
% \vspace*{4pt}
% \end{figure*}

\begin{Lemma}
\label{appr-2}
By selecting a reflection path resulting from a FR that contributes the most mutual information to the denominator of \eqref{eq-CRLB exact}, the CRLB for the FR-assisted localization can be approximated by:
\begin{equation}
\begin{split}
    \text{tr}\left ( \text{CRLB}(\bm{r}_{\text{RSS}}) \right ) \approx \sigma_{\text{RSS}}^2(\theta_{*}) \frac{4}{(L-1)}d_{*}^2, \\
\end{split}
\label{eq-RSS CRLB Approximate}
\end{equation}
where the subscript * denotes the selected FR, $L$ is the total number of FRs used for localization, $\sigma_{\text{RSS}}(\theta_{*})^2 = {\sigma_n^2}/{( \alpha + \frac{\sqrt{2}a}{a\theta_{*}+b} )^2}$, while $d_{*}$ and $\theta_{*}$ are the reflection distance and incident angle resulting from the selected FR, respectively. The tr(CRLB) distribution is then:
\begin{equation}
    P(\text{tr}\left ( \text{CRLB}(\bm{r}_{\text{RSS}}) \right ) \leq s) \approx F_{d_{*}} \left ( \frac{\sqrt{s(L-1)}}{2\sigma_{\text{RSS}}(\mathbb{E}\{\theta_{*}\})} \right ),
\label{eq-RSS CRLB distribution}
\end{equation}
where $s$ is stated localization accuracy; $\mathbb{E}\{\theta_{*}\}$ can be computed by using \cite[Proposition 2]{ho2026RSS}, while the CDF of $d_{*}$ is given by \cite[Proposition 1]{ho2026RSS}. The average CRLB can be obtained by using the CRLB distribution given in \eqref{eq-RSS CRLB distribution}.
\end{Lemma}

\textbf{Proposition \ref{prop-lower bound}} establishes a performance lower bound that approximates the optimal performance of the multi-FR-assisted localization system, while \textbf{Lemma \ref{appr-2}} provides a tractable expression for characterizing the achievable localization performance. Furthermore, by accounting for random FR placement, the overall localization performance can be evaluated through the distribution of the CRLB. Therefore, system designers can tune key system parameters, such as the RSS noise disturbance, path-loss exponent, and reflection gain, to gain insights into how the FR-assisted network are configured to meet the desired localization performance requirements.

\vspace{-0.4cm}

\section{Numerical Results}

\subsection{Simulation Setups}

We randomly generate $\mathcal{L}=10$ configurations and perform $\mathcal{M}=1,000$ Monte-Carlo (MC) runs for each configuration. The reflector centers and the target are randomly placed within the regions of $[-15,15]$ m$^2$. The inclination angles are randomly chosen from the region of $[0,\pi]$ rad. In addition, the measurement noise is assumed to be independent, such that the covariance matrix $\bm Q$ is set to $\bm Q = \sigma_n^2\mathbf{I}$, where $\sigma_n^2$ is the RSS noise power. Moreover, the performance of the maximum likelihood estimator (MLE) and the trace of CRLB are provided as a reference. The MLE is obtained by the MATLAB function ``lsqnonlin'' using the true values of the unknown parameters as the starting point.

\vspace{-0.4cm}

\begin{figure*}
  \begin{minipage}[b]{2\columnwidth}
  \centering
   \subfigure[]
  {\includegraphics[width=0.3\linewidth]{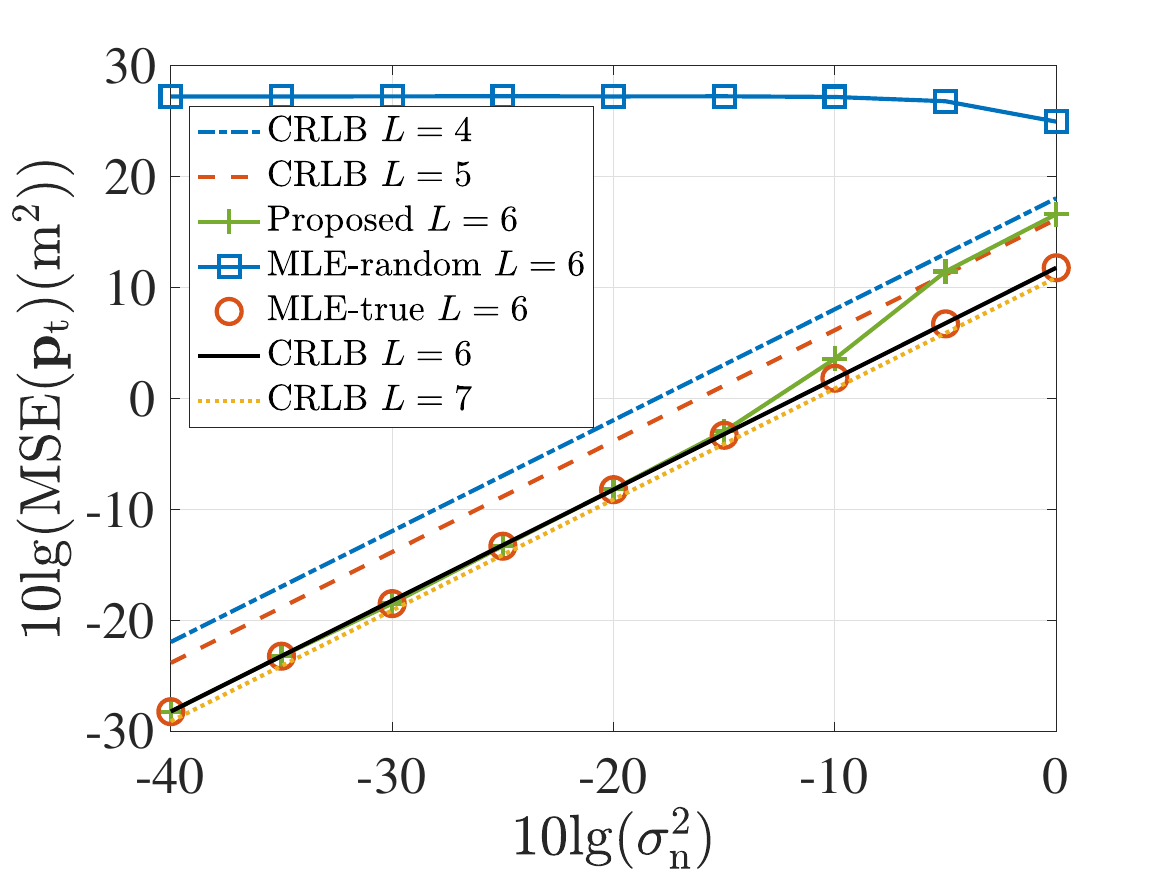}
  \label{MSEvsSigma}
  }
  \subfigure[]
  {\includegraphics[width=0.3\linewidth]{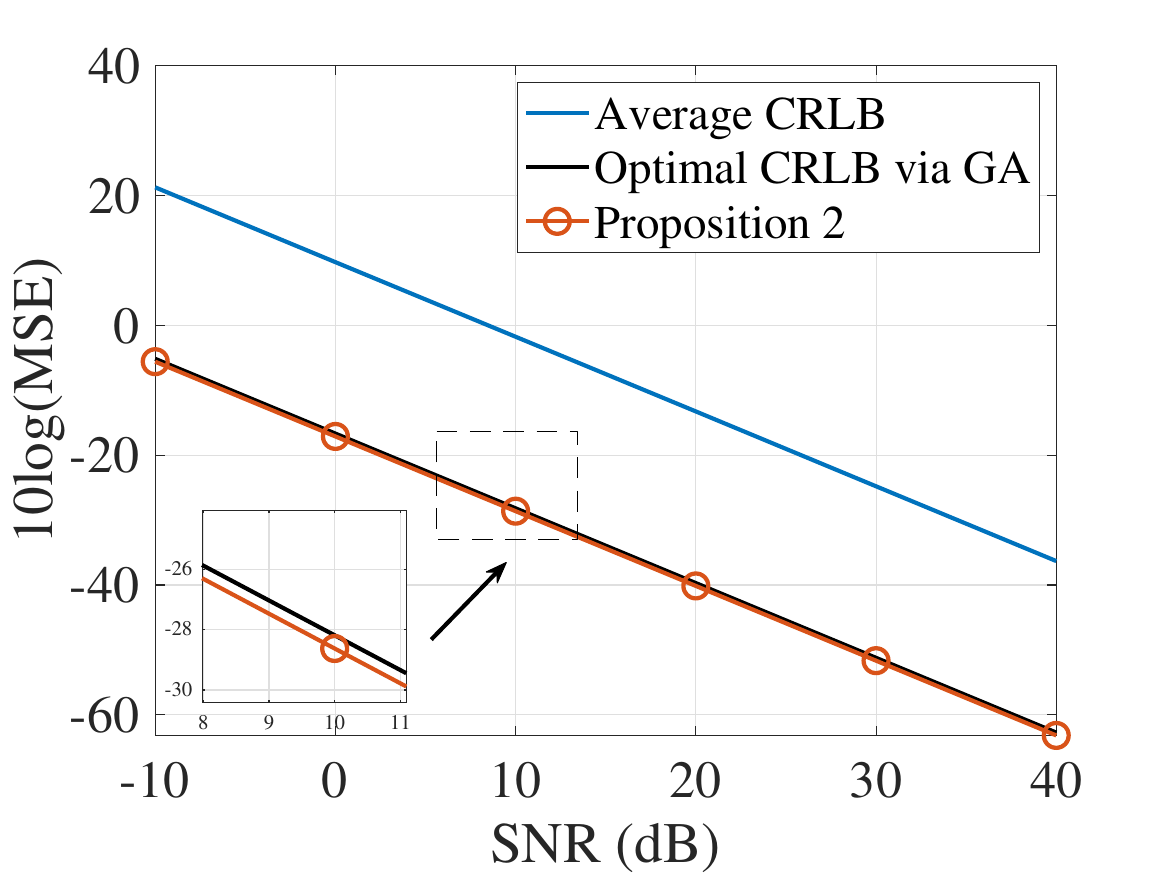}
  \label{fig-optimal_CRLB}
  }
  \subfigure[]
  {\includegraphics[width=0.3\linewidth]{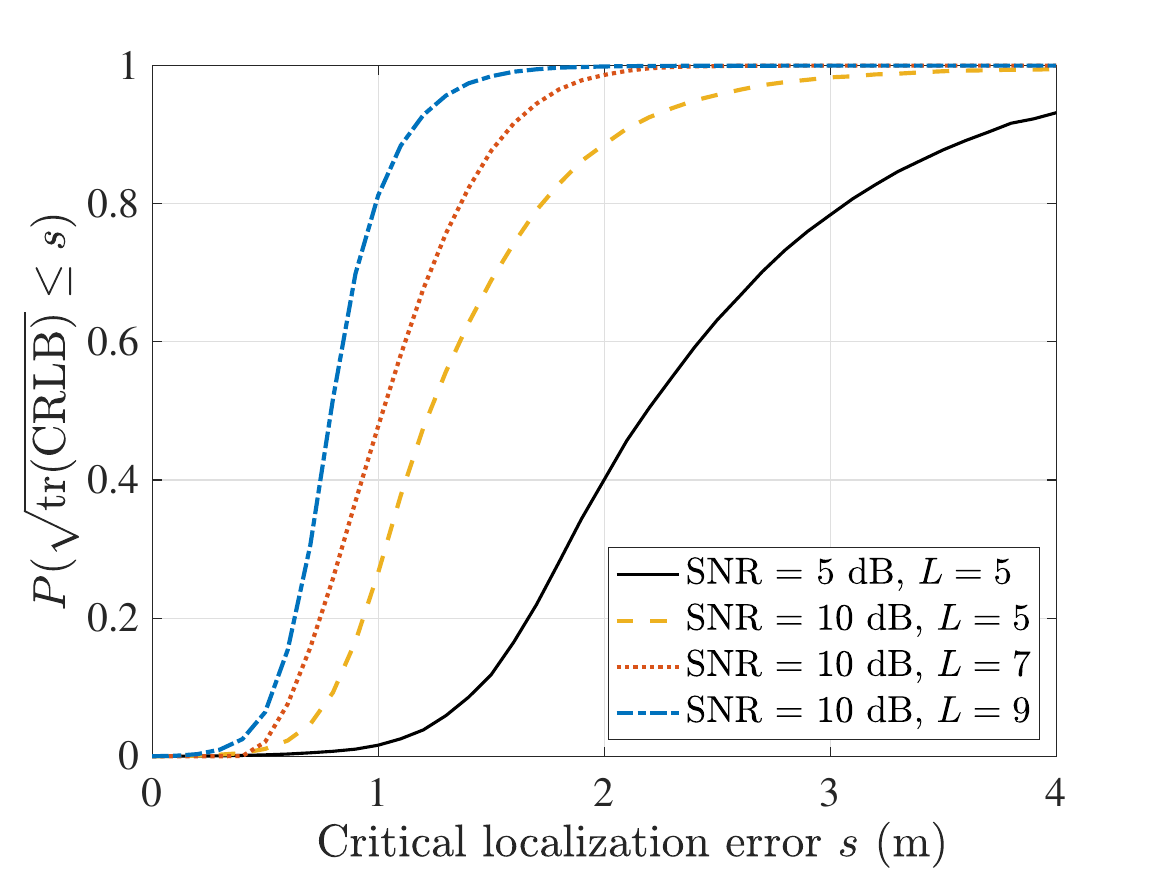}
  \label{fig-CRLB FR}
  }
  \caption{FR-aided localization: a) MSE performance comparison versus RSS measurement noise power when the number of FR is fixed at $L=6$; b) Optimal localization performance when $L = 5$; and c) CRLB distribution under different SNR and FR number.}
  \label{fig-overall}
  \end{minipage}
\vspace{-0.4cm}
\end{figure*}

\subsection{Results}

\underline{\textit{Performance of FR-aided Localization}}: Figure \ref{MSEvsSigma} examines the performance of the proposed solution as the RSS measurement noise power varies when the number of reflectors is fixed at $L=6$. The results are shown in Fig. \ref{MSEvsSigma}. For comparison, the CRLB and MLE with true and random initialization are included. It is observed from Fig. \ref{MSEvsSigma} that the proposed solution closely attains the CRLB when the RSS measurement noise power is not greater than $\sigma_n^2=0.1$. The ``MLE-true'' can always reach the CRLB owing to an ideal initialization, which is unavailable in practice. To examine the influence of the initial value, we also include the curve ``MLE-random'', which plots the performance of MLE with random initialization. It is seen that, with an unreliable starting point, the MLE performance becomes poor, owing to the local convergence or even divergence. By contrast, our approach does not require any initialization. Figure \ref{MSEvsSigma} also plots the CRLB curves of different numbers of reflectors. Since localization performance improves with the deployed FR number, deploying additional FRs can further enhance localization accuracy.

\underline{\textit{Optimal Localization Performance}}: Next, we investigate the optimized reflector placement with $5$ reflectors using the GA and validate the lower bound given by \textbf{Proposition \ref{prop-lower bound}} under different SNR values. We employ the GA to obtain the optimized placement of $5$ reflectors and subsequently compute the associated value of tr(CRLB). The results in Fig. \ref{fig-optimal_CRLB} support that the benchmark produced by \textbf{Proposition \ref{prop-lower bound}} can serve as a tight lower bound to indicate the optimized localization performance without requiring complicated and lengthy computation, as the GA. The localization performance averaged over $10^5$ realizations with the reflectors randomly placed as described in the simulation setup is shown in Fig. \ref{fig-optimal_CRLB}, together with the optimized values. It illustrates that the CRLB obtained from the optimized placement of the reflectors is lower than the average CRLB by nearly $15$ dB. Thus, it is vital to optimize the network configuration to achieve the stated localization requirements.

\underline{\textit{Overall Localization in Random FR Placements}}: To characterize the overall localization performance of the FR-aided localization scheme, we evaluate the CRLB in Proposition under random FR deployment. Specifically, the FR center points are modeled as a HPPP with density $\lambda_r = 1 \times 10^3$, while the length and inclination angle of each FR are uniformly distributed over $[5,15]$ m and $[0,\pi]$ rad, respectively. The path-loss exponent is set to $\alpha = 4.3$, while the reflection coefficients are given by $a = 0.573$ and $b = 0.1$. Figure \ref{fig-CRLB FR} shows that the localization performance improves with the number of FRs used for localization. Nevertheless, the performance gain achieved by increasing the FR density gradually saturates as the FR number becomes sufficiently large. Hence, an appropriate FR number should be selected to balance the tradeoff between deployment cost and the required localization accuracy.

\vspace{-0.4cm}

\section{Conclusion}

This paper investigated FR-assisted target localization in scenarios where the direct path between the target and the deployed Rx is not available. A FR-assisted localization problem was formulated, and a low-complexity estimator was developed for high-precision target localization. To characterize the achievable performance, the CRLB was derived and further adopted to optimize the placement and orientation of the FRs. Moreover, the distribution of the CRLB under random FR deployment was analyzed to assess how network configurations affect the overall localization performance. Simulation results demonstrated that our scheme closely approaches the CRLB accuracy, while the developed analytical results provide useful guidelines for FR deployment and network design.

\vspace{-0.2cm}

\appendices

% \section{}
% \label{appendix-appr crb}

% It is noticed that $\frac{\partial \bar{\theta}}{\partial x_t}$ in \eqref{d-loss J} can be written as: $\frac{\partial \bar{\theta}}{\partial x_t} \leq \frac{(x_t - x_l)}{d_l^2}$. Thus, we have $\frac{\partial 2\ln(a\theta_l +b)}{\partial x_t} 
% \approx -\frac{2a}{a\theta_l+b}\cdot\frac{(x_t - x_l)}{d_l^2}$. By making use of the Cauchy-Schwarz inequality and Taylor expansion, it is also realized that the incident angle can be approximated by: $\theta_l \geq \frac{\sqrt{2}}{2} \frac{R}{d_l}$. As a result, the approximate CRLB can be rewritten using the updated $\frac{\partial \bar{\theta}}{\partial x_t}$ and $\theta_l$. %Thus, the proof is complete.

\bibliographystyle{ieeetr}
\bibliography{ref}

\end{document}